# Surgical data science for safe cholecystectomy: a protocol for segmentation of hepatocystic anatomy and assessment of the critical view of safety

Pietro Mascagni[1, 2, 3†], M.D.; Deepak Alapatt[1†], M.Sc.; Alain Garcia[3], M.D.; Nariaki Okamoto[4], M.D.; Armine Vardazaryan[1], M.Sc.; Guido Costamagna[2], M.D., Ph.D.; Bernard Dallemagne[3, 4], M.D.; Nicolas Padoy[1, 3], Ph.D.

1. ICube, University of Strasbourg, CNRS, France
2. Fondazione Policlinico Universitario A. Gemelli IRCCS, Rome, Italy
3. IHU-Strasbourg, Institute of Image-Guided Surgery, Strasbourg, France
4. Institute for Research against Digestive Cancer (IRCAD), Strasbourg, France

Minimally invasive image-guided surgery heavily relies on vision. Deep learning models for surgical video analysis could therefore support visual tasks such as assessing the critical view of safety (CVS) in laparoscopic cholecystectomy (LC), potentially contributing to surgical safety and efficiency. However, the performance, reliability and reproducibility of such models are deeply dependent on the quality of data and annotations used in their development. Here, we present a protocol, checklists, visual examples and a tool to promote consistent annotation of hepatocystic anatomy and CVS criteria. We believe that sharing annotation guidelines can help build trustworthy multicentric datasets for assessing generalizability of performance, thus accelerating the clinical translation of deep learning models for surgical video analysis.

† Pietro Mascagni and Deepak Alapatt contributed equally and share first co-authorship.
Correspondence to: Pietro Mascagni, M.D., p.mascagni@unistra.fr
This work was partially supported by French State Funds managed by the "Agence Nationale de la Recherche (ANR)" through the "Investissements d'Avenir" (Investments for the Future) Program under Grant ANR-10-IAHU-02 (IHU-Strasbourg) and through the National AI Chair program under Grant ANR-20-CHIA-0029-01 (Chair AI4ORSafety).

# Index

## 1. Vision: the double-edged sword of modern surgery

Image-guided procedures such as laparoscopic, endoscopic, and radiological interventions heavily rely on imaging to guide gestures (1). Such minimally invasive approaches have shown significant value for patients and healthcare systems. However, the subconscious assumptions typical of human visual perception can lead to erroneous interpretations of endoscopic and radiological images, contributing to serious adverse events that were rarer during traditional, open surgery procedures (2).

The Shepard elephant depicted in Figure 1 is a classic example of this type of optical illusion (3).

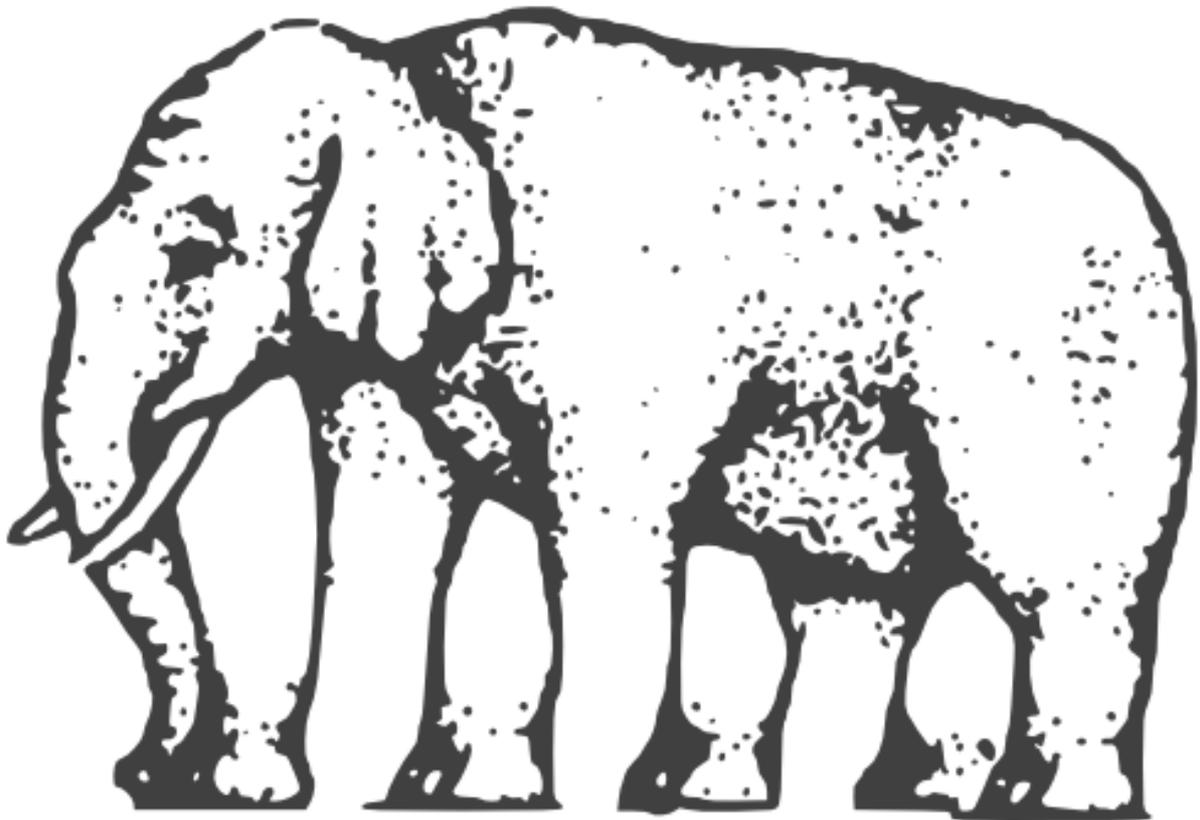

**Figure 1. Shepard elephant or the impossible elephant.** This is considered an impossible object, a type of optical illusion derived from the subconscious projection of a two-dimensional object in the three-dimensional space performed instantly by the visual system.

## 2. Deep learning for safe cholecystectomy

The problem of bile duct injuries (BDIs) in laparoscopic cholecystectomy (LC) well exemplifies the case of a serious adverse event caused by visual illusion. In fact, it was found that 97% of BDIs happen due to visual perceptual illusions (2). In the classical BDI, the surgical operator believes to be clipping and cutting the cystic duct while actually dividing the common bile duct. To prevent the visual illusion causing BDI, in 1995 Strasberg et al (4) proposed to perform the critical view of safety (CVS) technique to conclusively identify the cystic duct. Today, this safety view has been so extensively studied that all guidelines on safe LC recommended obtaining CVS prior to the division of the cystic duct (5-7). However, CVS assessment is qualitative and subject to observer interpretation (8-10). This might explain the non-decreasing rate of BDI (11), remaining at least 3 times more common in LC than open cholecystectomy despite 3 decades of CVS and efforts from the surgical community (12).

Surgical data science, a multidisciplinary field using the vast amount of digital data generated in surgery to enhance care (13), could help in this regard. In fact, deep learning models for surgical video analysis can be trained to guide the surgeons towards safe areas of dissection (14), reliably segment hepatocystic anatomy and unequivocally assess CVS (15) (Figure 2), and produce selective video documentation of this safety step of LC (16). Once thoroughly validated in their performance and ability to generalize, such models could be used intraoperatively to enhance surgeons awareness, potentially contributing towards surgical safety.

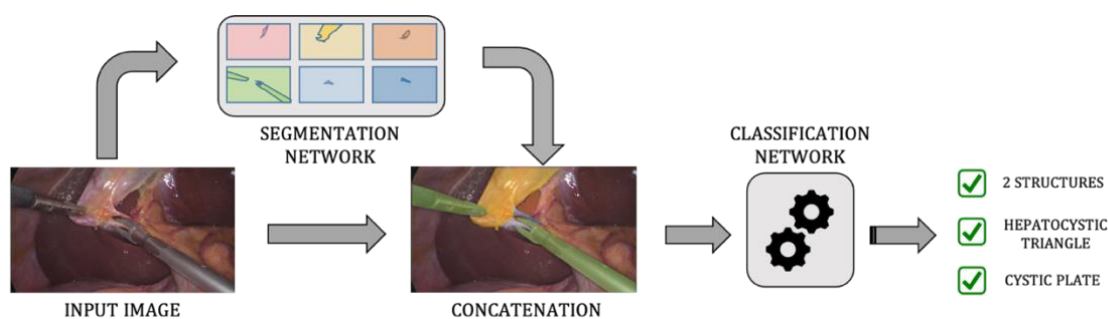

**Figure 2. Schematic representation of DeepCVS** (15). This 2-stage model analyses laparoscopic images by first segmenting hepatocystic anatomy and surgical tools and then passing this information to a network tasked with classifying each of the 3 CVS criteria as achieved or not.

## 3. Need for shared annotation protocols

Deep learning is notably highly dependent on the quantity and quality of data. In surgery, large and well annotated datasets are scarce if not absent, limiting research and development efforts despite the widely recognized potential of computer assistance (17).

The sensitive and private nature of surgical data is certainly a factor contributing to such paucity of publicly available datasets. However, processes to collect and share clinical data exists (17).

We believe a significant bottleneck for the generation of meaningful surgical datasets resides in annotation. Annotating surgical data with ground truth, the gold standard information used to train and assess learning models, can be a difficult and costly process (18). This is especially true when annotating instances of high surgical semantics such as CVS in LC, requiring the development of methods for consistent assessments from overburdened healthcare professionals (10, 19).

Sharing annotation protocols would accelerate the construction and allow the amalgamation of annotated datasets from multiple centres, greatly contributing to the development of robust and reliable deep learning models. Furthermore, published annotation protocols would facilitate the development of external, annotated datasets. These datasets could be used to independently test models, thereby mitigating the need for *a posteriori* analysis of models outputs (20).

Finally, publishing annotation protocols would also help gain better insights and discuss the clinical relevance of models performance. In fact, by inspecting how annotations were defined, one could appraise the clinical soundness of the ground truth used to compute models accuracy. In brief, publishing annotation protocols could contribute towards:

- Reproducibility, by providing guidelines for collaborators and researchers wanting to perform an independent assessment of models
- Trust, by offering insights on clinical relevance and reliability of the ground truth used for training and testing deep learning models

Overall, we believe public and shared annotation protocols could help accelerate the clinical assessment and uptake of deep learning models for surgical video analysis.

## 4. Annotation protocol and workflow

The annotation process described below includes the collection of surgical videos, the video assessment of CVS criteria, the extraction of video-frames and, finally, the segmentation of hepatocystic anatomy and annotation of CVS criteria on the extracted images. The whole process is illustrated in Figure 3.

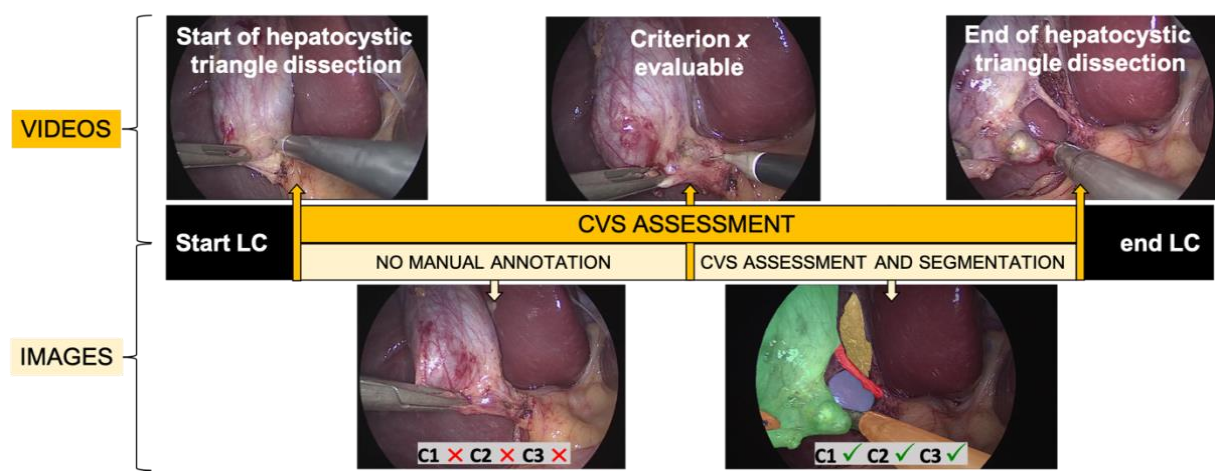

**Figure 3. Schematic representation of the annotation process.** The top of the figure shows the visual cues marking the first incision on the hepatocystic triangle, a criterion becoming evaluable (in this case, the cystic duct and artery start to be visible in their tubular appearance) and the first clip applied on the cystic duct, the 3 timestamps defining the region of interest for video assessment of CVS and video frames extraction. The bottom of the figure represents the extraction of key frames for manual segmentation of hepatocystic anatomy and CVS criteria annotation.

### 4.1 Collecting LC videos for CVS assessment

Endoscopic videos of LC procedures are the primary source of data used for hepatocystic anatomy segmentation and CVS assessment. Consecutive and unselected LC videos are collected to generate a representative and diverse dataset, including videos of procedures with anatomical variations and varying degree of inflammation. However, videos showing one of the following deviations from the normal procedural workflow induced by failure to achieve CVS were excluded: fundus first or anterograde cholecystectomy, subtotal or partial cholecystectomy, intraoperative cholangiogram, conversion to open surgery and abortion of the LC procedure.

### 4.2 Defining a region of interest for CVS assessment

Surgically, CVS should be achieved prior to the division of the cystic duct and artery by careful dissection of the hepatocystic triangle. It follows that the region of interest for CVS assessment should start with the hepatocystic triangle dissection and end when the first clip is applied on the cystic duct or the cystic artery (Figure 3). This region of interest of the video is used for training and testing deep learning models for hepatocystic anatomy segmentation and CVS criteria assessment.

### 4.3 Video annotation

#### 4.3.1 Assessing CVS criteria in videos

According to Strasberg et al (4, 21), CVS is defined as the view of 2 tubular structures, the cystic duct and the cystic artery, connected to the gallbladder (2-structure criterion, C1), a hepatocystic triangle cleared from fat and connective tissues (hepatocystic triangle criterion, C2) and the lower part of the gallbladder separated from the liver bed (cystic plate criterion, C3). For consistency, each of these 3 CVS criteria should be assessed as either achieved or not, in a binary fashion; if 3 out of 3 criteria are achieved, then CVS is obtained (10).

### 4.4 Image annotation

#### 4.4.1 Extracting key CVS images for manual annotation

To populate the dataset, CVS images are extracted from videos and manually annotated. A video-frame extraction process was designed to minimize selection bias and to find a good trade-off between the annotation effort and having a representative and sizeable dataset. This process entails the following steps (Figure 3):

1. The region of interest of the video is reviewed to annotate the first moment (i.e., timestamp) when any of the 3 CVS criteria is considered evaluable (Criterion $x$ evaluable).

    i. A criterion is considered evaluable when its anatomical elements can be seen but the criterion is not achieved, e.g, part of the cystic duct is visible but 2 tubular structures are not clearly visible => C1 evaluable. Note, it is important to be conservative with the "Criterion $x$ evaluable" timestamp (i.e., place the timestamp as soon as elements of a criterion start to be visible) to avoid missing images that should be manually annotated.

2. Video-frames between the start of the region of interest for CVS assessment and the "Criterion $x$ evaluable" timestamp are automatically marked as having no criteria achieved.

3. Temporally regularly spaced key video-frames between the "Criterion $x$ evaluable" timestamp and the end of the region of interest are manually annotated.

    i. If no "Criterion x evaluable" timestamp occurs during the procedure, temporally regularly spaced key frames from the entire region of interest are manually annotated.

4.4.2 Assessing CVS criteria in images

CVS is more difficult to assess in static images, with several investigations reporting a lower inter-rater agreement in photo assessment compared to video assessment of CVS (10, 22, 23).

A checklist to improve consistency of CVS criteria assessments during annotation and revision of images is presented at the end of the manuscript.

4.4.3 Segmenting hepatocystic anatomy in images

The following instances are pixel-wise segmented due to their relevance to CVS:

- Gallbladder: The pear-shaped organ excised during LC.

- Cystic duct: The tubular-shaped biliary structure divided during LC.

- Cystic artery: The tubular-shaped vascular structure divided during LC.

- Cystic plate: The connective tissue exposed separating the gallbladder from the liver bed.

- Hepatocystic triangle dissection: The dissected "windows" in the hepatocystic triangle exposed during LC.

- Surgical instruments: The laparoscopic tools used during LC.

- Ignore: Label used to mark anatomical variations.

- Background: An implicitly segmented class corresponding to every pixel not corresponding to one of the above-mentioned classes.

Rules of thumb to improve segmentation consistency is presented at the end of the manuscript.

### 4.5. Notes on annotation

4.5.1 Annotators

Assessing CVS criteria entails evaluating the quality of the surgical dissection, requiring annotators with domain knowledge. For this reason, annotators of CVS criteria should be surgeons specifically trained on the principles of a safe cholecystectomy (24). In addition, three or more surgeons should independently assess each video/image to allow the deep learning model to learn the most commonly held interpretation of CVS rather than a single assessment. Finally, disagreements in CVS criteria assessments should not be mediated to avoid introducing biases.

Gross hepatocystic anatomy can be more objectively defined, especially when assessed on videos, outside of the time-constraints and stress of operating rooms. Furthermore, contouring variability (i.e., including or excluding a few pixels around an anatomical structure) should not have clinical implications in our use-case. For this reason, semantic segmentation of surgical images is performed by computer scientists trained on the fundamentals of hepato-biliary anatomy and surgeons alike. Given the time required to extensively segment surgical images, each extracted frame is semantically segmented once. However, all semantic segmentations are revised by at least one independent annotator to improve consistency.

### 4.5.2 Annotation process

Before starting the annotation process, an annotation protocol, checklists, and flashcards should be shared with enrolled annotators to collect feedback and discuss improvements.

During the annotation process, regular meetings should be held with all the involved annotators to give feedback based on randomized and anonymized reviews of annotations to rule out major inconsistencies with the provided definitions. Such reviews are qualitative or make use of standard inter-rater agreements metrics such as Cohen's kappa.

To promote consistency, annotators should watch videos before annotating images and should return to videos in case contextual information is needed to conclusively identify anatomical structures present in the frame being annotated. Finally, annotators should review all images extracted from a single procedure sequentially to spot and correct eventual annotation inconsistencies.

### 4.5.3 Annotation tools

In our experience, annotating LC videos with timestamps and an assessment of CVS criteria can be done using any compatible media player and word editor.

On the other hand, to efficiently annotate a large number of images with CVS criteria and segmentation we required specific annotation software.

A custom software was developed for effective annotation of CVS criteria in images (25). The custom CVS annotation software developed at [CAMMA](#) (Computational Analysis and Modeling of Medical Activities research group, ICube, University of Strasbourg, IHU Strasbourg, France) allows to quickly navigate between images, label CVS criteria and extra information with forms or free text comments and open videos at the exact time point showing the image to annotate for quick review. The software can be downloaded at https://github.com/CAMMA-public/cvs_annotator.

Finally, LabelMe (26), a freely available, open-source software was used for polygon-based semantic segmentation of surgical images. The source code of LabelMe was also modified to allow opening videos at the exact time point showing the image to annotate.

## 5. Conclusion

Here, we stress the need for strictly defining and sharing guidelines to annotate surgical images and present our protocol for consistent hepatocystic anatomy segmentation and CVS criteria assessment. We hope that this effort will set a precedent and facilitate a discussion with clinicians and computer scientists alike on the clinical value and technical challenges associated with developing meaningful surgical datasets. The development of such datasets could strongly accelerate the translation of deep learning models to enhance safety and efficiency of surgical care.

## 6. CVS criteria checklist

| C1 – 2-structure criterion | |
|---|---|
| **Definition:** The cystic duct and the cystic artery are the only 2 tubular structures connected to the gallbladder | |
| **Not achieved:**<br>  a) ≠2 tubular structures<br>  b) <u>Not tubular-appearing</u> due to incomplete dissection, point of view, instrument occlusion, etc.<br>  c) Not seen <u>entering the gallbladder</u> due to incomplete dissection, point of view, instrument occlusion, etc. | ✗ |
| **Achieved:**<br>  b) View of <u>2 tubular structures connected</u> to the gallbladder<br>      I. Note: The criterion is considered achieved even if the presence of an extra tubular structure cannot be excluded due to insufficient dissection (i.e., non-achievement of C2 and/or C3) | ✓ |
| **C2 – Hepatocystic triangle criterion** | |
| **Definition:** The hepatocystic triangle is cleared of fat and/or connective tissue so that an unimpeded view of the hepatocystic triangle is obtained | |
| **Not achieved:**<br>  a) <u>No full-thickness</u> dissection (i.e., the underlying anatomy is not visible through the dissected windows)<br>  b) <u>Inability to confirm/exclude the presence of other structures</u> running between the cystic duct and the cystic artery and between the cystic artery and the cystic plate | ✗ |
| **Achieved:**<br>  c) The presence of 2 <u>full-thickness dissected windows</u> in the hepatocystic triangle (between the cystic duct and the cystic artery, between the cystic artery and the cystic plate) is verified through vision or interposition of an instrument<br>      I. Note: In cases with a single tubular structure connected to the gallbladder (i.e., no cystic artery), there will not be 2 windows, but the criterion can still be achieved | ✓ |
| **C3 – Cystic plate criterion** | |
| Definition: The lower part of the gallbladder is divided from the liver bed to expose the cystic plate | |
| **Not achieved:**<br>  a) The cystic plate is <u>not visible along the whole inferior margin of the gallbladder</u> due to incomplete anterior or posterior dissection, instrument or blood occluding the view, etc.<br>  b) <u>Inability to confirm/exclude the presence of other structures</u> running on the cystic plate | ✗ |
| **Achieved:**<br>  c) The cystic plate is visible below the <u>whole inferior margin</u> (i.e., from anterior to posterior view) of the gallbladder<br>      I. Note: Should resemble a "hanging gallbladder" in open cholecystectomy<br>      II. Note: A fatty/thick cystic plate and a dissection between the subserosa inner and subserosa outer layers (i.e., dissection close to the gallbladder) does not prevent the criterion from being achieved | ✓ |

Note, redundancy between the "Not achieved" and "Achieved" rows is purposefully included to reiterate and explicitly elaborate the criteria definitions for annotators.

Annotation Protocol for Safe Cholecystectomy    13## 7. Segmentation rules of thumb

| Segment only well dissected structures |
|---|
| a) Only segment the tubular appearing, texture-specific cystic duct and artery |
| b) If a window of the hepatocystic triangle is only partially dissected (i.e., the background is not visible), don't segment it |
| **Segment only what is visible** |
| c) When any labelled class is visible through the hepatocystic triangle (e.g., gallbladder, cystic plate in tangential views, etc.), label the visible class instead of the dissected hepatocystic triangle |
| d) When margins are not clearly visible due to poor lighting, adjust the image brightness |
| **Anatomical margins** |
| e) To define the lower margin of the gallbladder and the distal margins of the cystic duct and the cystic artery, follow a straight line that passes through the intersection of the gallbladder with the cystic duct and the cystic artery |
| **Anatomical variations** |
| f) Segment the cystic duct and the cystic artery only if you are 100% sure of their correct identification. Use the video to confirm anatomy |
| g) If the third tubular structure arises from a bifurcation of what you have conclusively identified as the cystic artery, then segment both bifurcation branches as cystic artery |
| h) When you cannot conclusively identify a tubular structure, segment it with the "Ignore" label |
| **Surgical instruments** |
| i) The anatomy or a separate tool visible through the fenestrated jaws of instrument (e.g., fenestrated grasper, bipolar) should not be segmented. If a portion of the same tool is visible through the fenestrated jaws, then segment it |

## 8. CVS criteria annotation flashcards

Here, we present and describe a set of laparoscopic images labelled with an assessment of CVS criteria in order to illustrate and explicit the rationale behind CVS annotation.

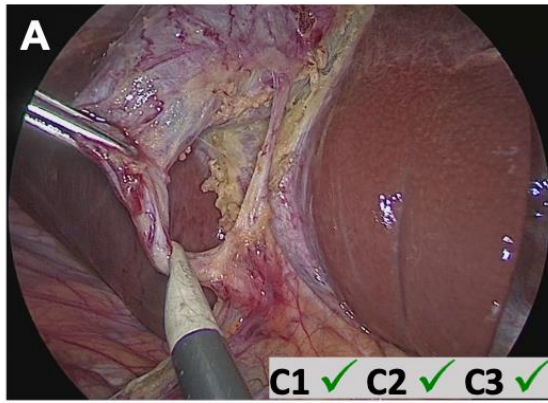

The view of the hepatocystic triangle is unimpeded, the cystic plate is visible along the whole inferior margin of the gallbladder and two tubular structures are visible entering the gallbladder (anterior, medial view)

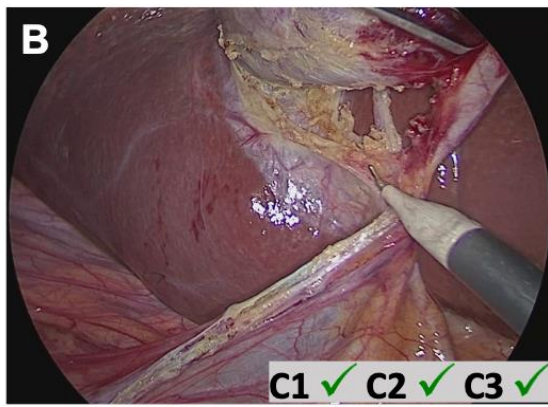

The view of the hepatocystic triangle is unimpeded, the cystic plate is visible along the whole inferior margin of the gallbladder and two tubular structures are visible entering the gallbladder (posterior, lateral view)

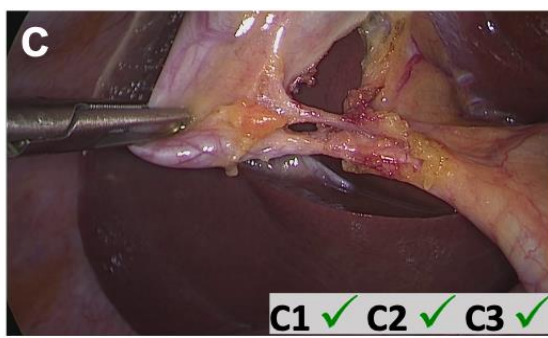

The cystic plate is visible along the whole inferior margin of the gallbladder, two tubular structures are clearly visible and an extensive dissection close to the gallbladder allows to exclude the presence of other structures

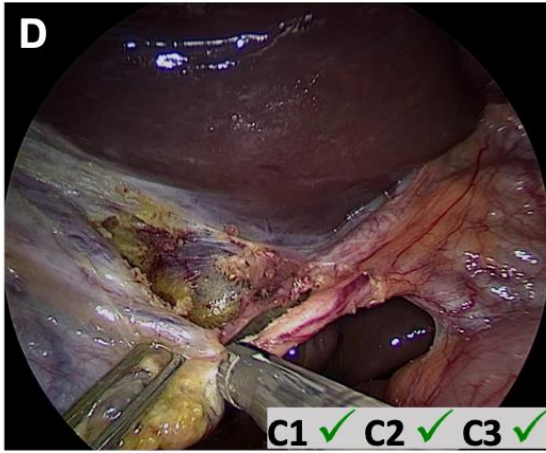

Despite the rotated viewing axis, the presence of extra structures in the hepatocystic triangle or on the cystic plate can be excluded and two tubular structures are visible entering the gallbladder

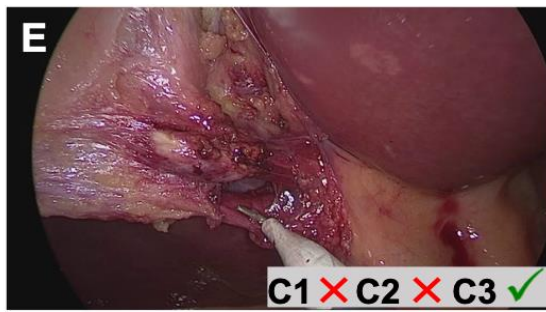

The cystic plate is clearly visible, but the undissected tissue around the cystic artery impedes the view of two tubular structures and prevents from excluding the presence of other structures in the hepatocystic triangle

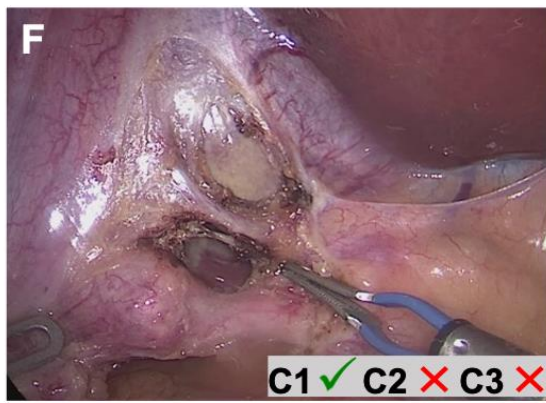

Two tubular structures are visible, but the presence of extra structures in the hepatocystic triangle cannot be excluded and the cystic plate is not visible below the whole inferior margin of the gallbladder

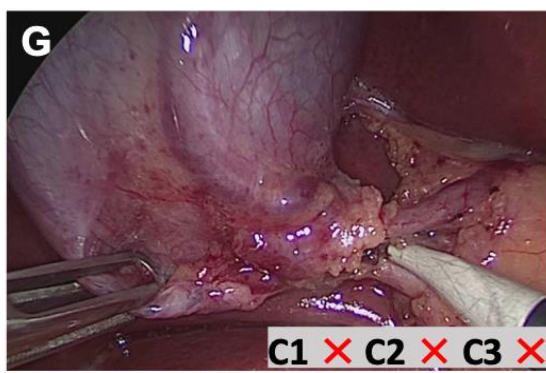

The cystic duct and artery are not visible in their tubular appearance due to the retraction angle and a surgical instrument occluding the view. Neither the dissection between the cystic duct and artery nor the cystic plate are visible

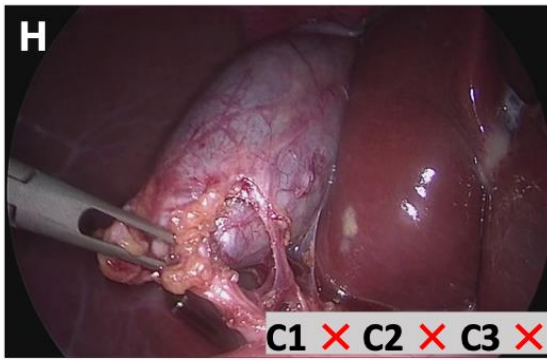

A horizontal structure is visible between the cystic duct and artery, the dissection in the hepatocystic triangle is incomplete and the cystic plate is not visible

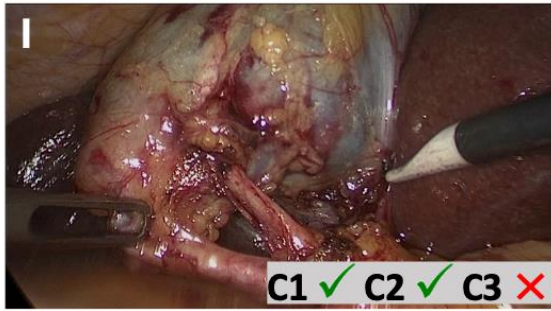

The full-thickness dissection of the hepatocystic triangle allows to clearly see only two tubular structures but the cystic plate is not visible

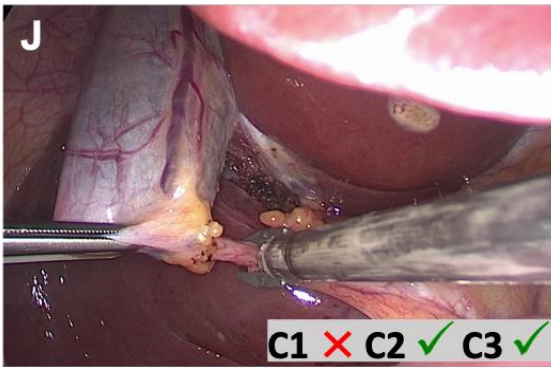

The cystic plate is visible along the whole inferior margin of the gallbladder and the hepatocystic triangle is well dissected, but only one tubular structure is present

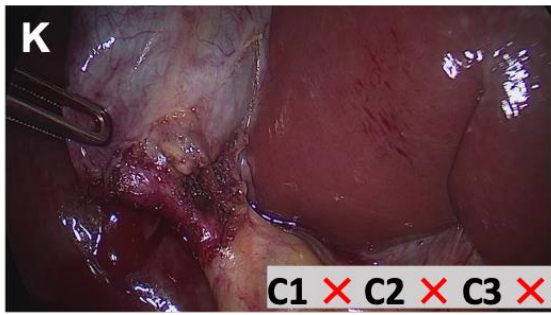

Tangential point of view limits assessment

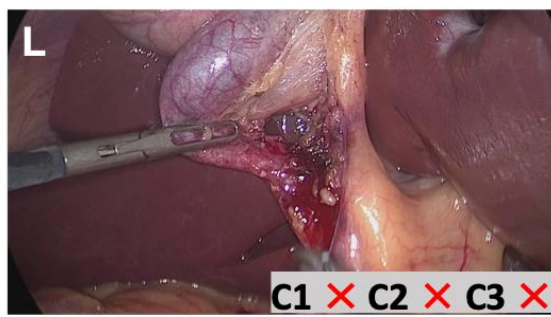

Undissected tissue in the hepatocystic triangle prevents the view of a second tubular structure and the cystic plate

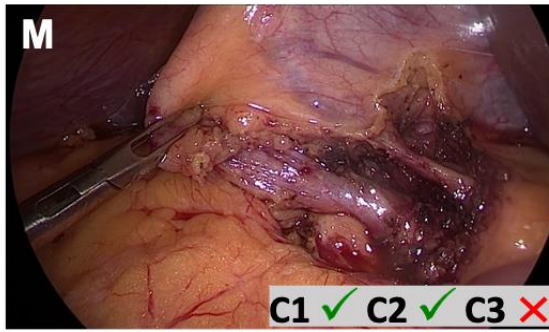

Two tubular structures and the full-thickness dissection of the hepatocystic triangle are visible, but the presence of extra structure on the cystic plate cannot be excluded due to limited dissection

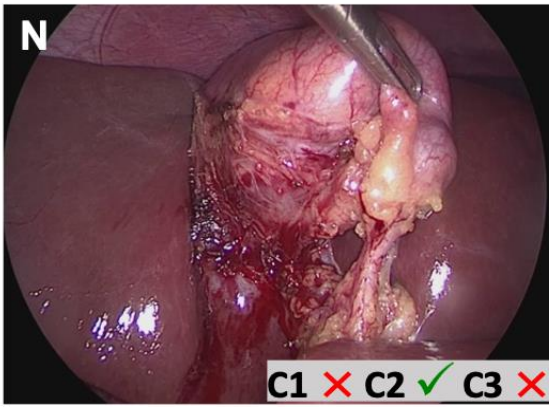

The vision of the cystic plate is partially occluded by blood and only one tubular structure is visible despite the full-thickness dissection of the hepatocystic triangle

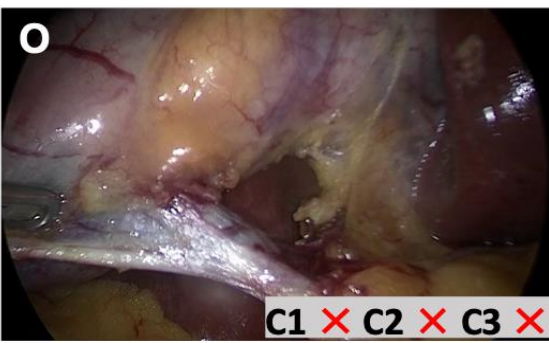

The limited full-thickness dissection of the hepatocystic triangle shows only one tubular structure and does not allow excluding the presence of extra structures on the cystic plate

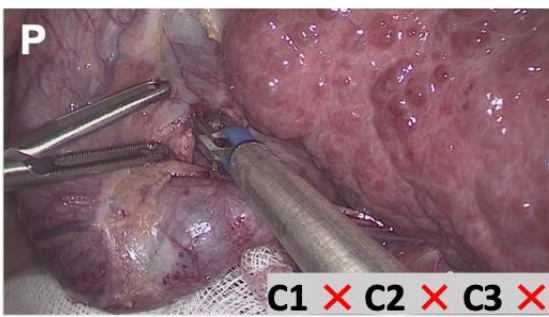

A bipolar grasper occludes the view of the hepatocystic triangle

## 9. Hepatocystic anatomy segmentation flashcards

Here, we present a few illustrations of manual segmentation to visualize rules for annotation of hepatocystic anatomy and surgical tools.

- Segmentation legend

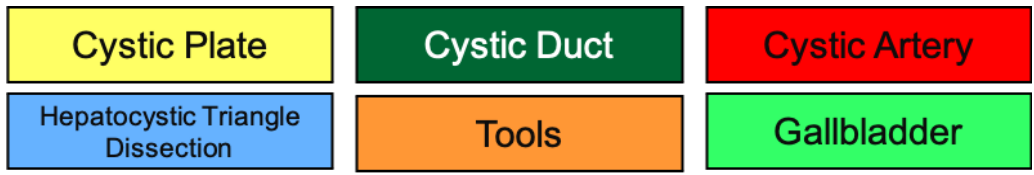

- Complete segmentation of hepatocystic anatomy in a CVS achieved case

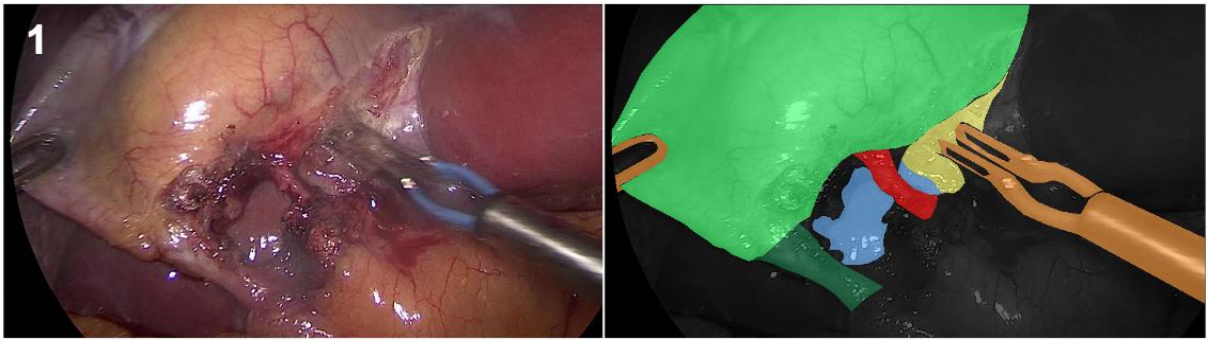

- Segment any extra tissue sitting on anatomical structures

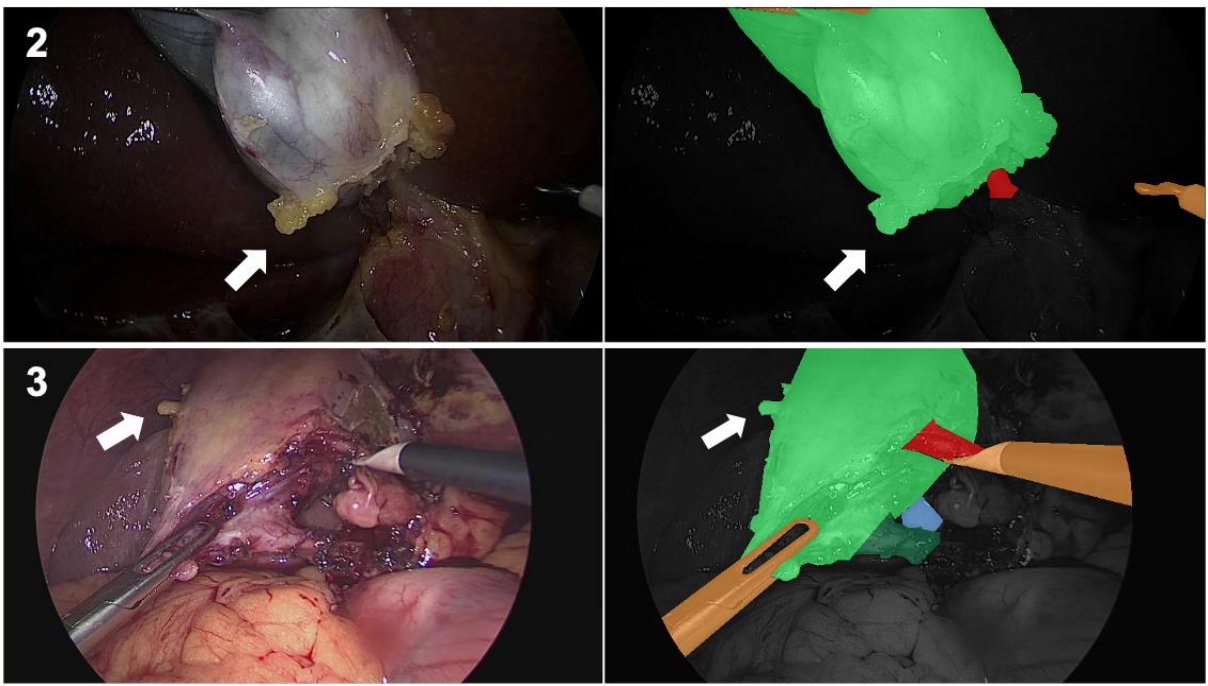

- Unless texture specific transitions are visible, use the inferior margin of the gallbladder to draw a border between the gallbladder and cystic duct/cystic artery

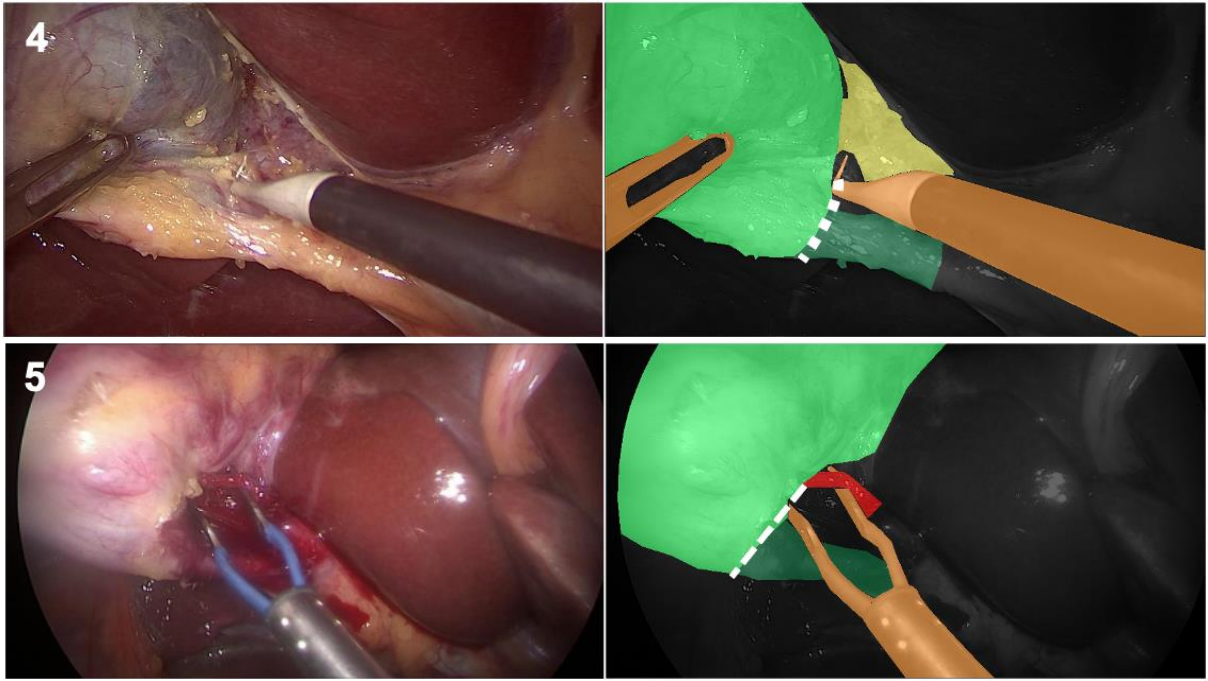

- Adjust image brightness if needed to better identify margins of structures

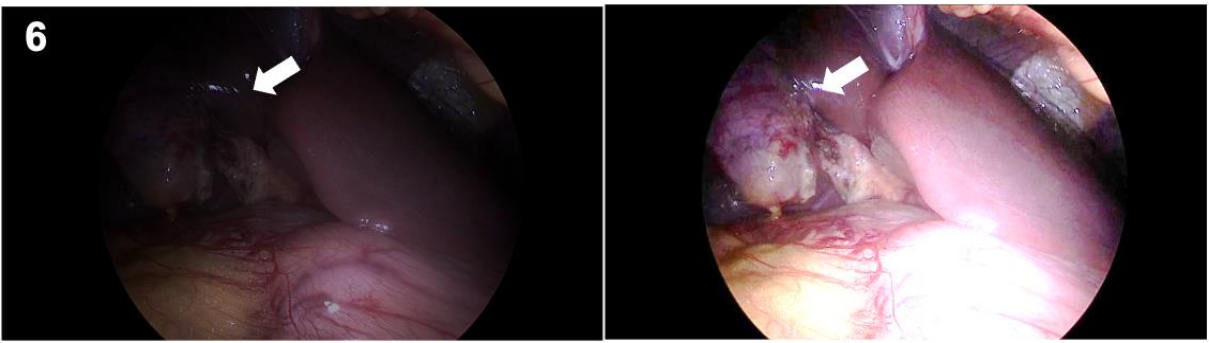

Annotation Protocol for Safe Cholecystectomy 20

- Use the video only to identify and delineate structures that are visible in the frame you are annotating

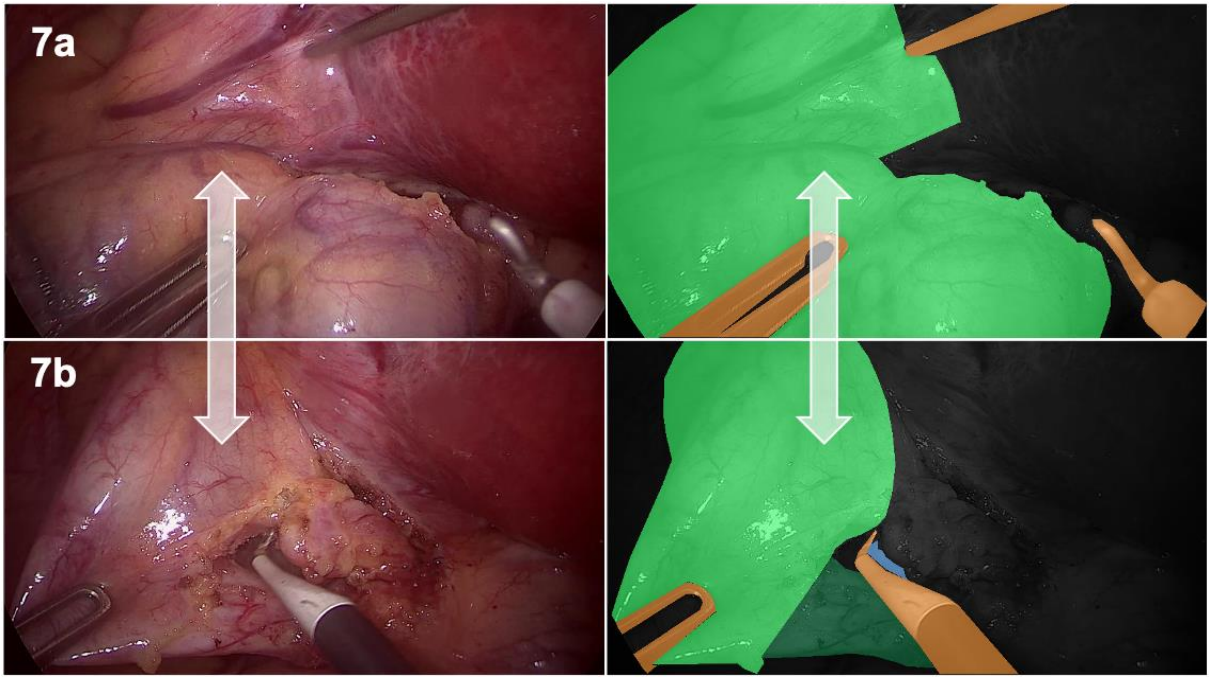

- Label any relevant anatomy or tools visible through or occluding the hepatocystic triangle dissection

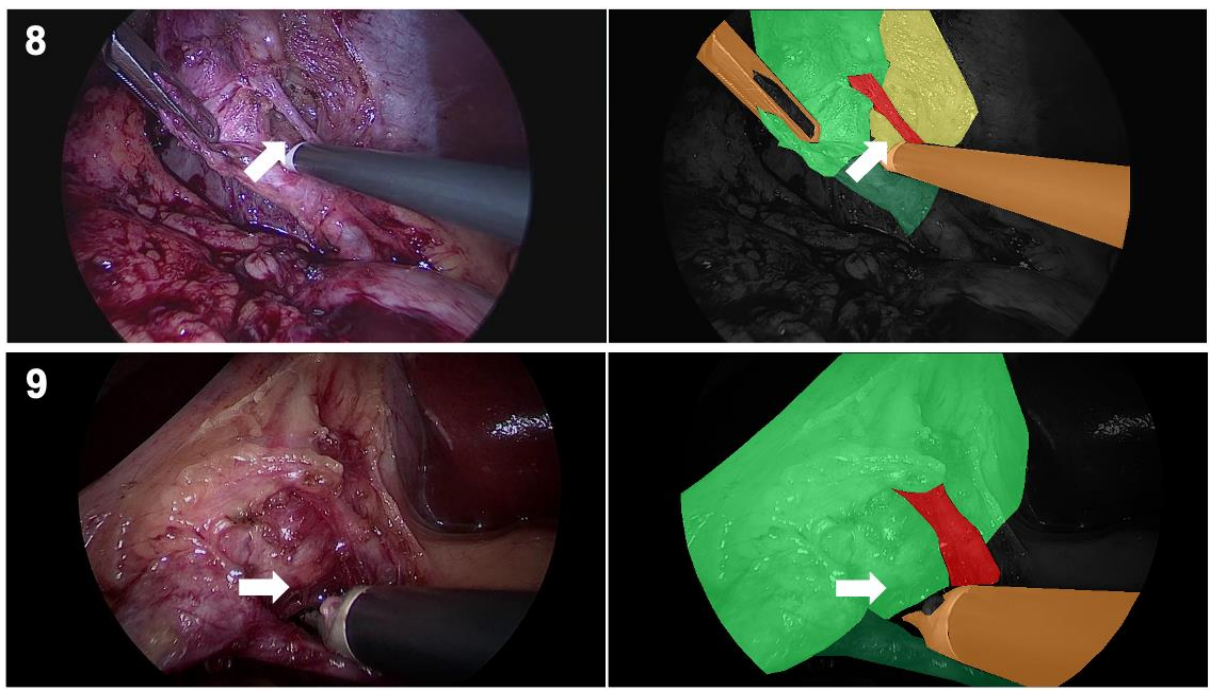

- Label the region visible through the fenestrated jaws of instruments such as grasper and bipolar as background unless a portion of the same instrument is visible through it

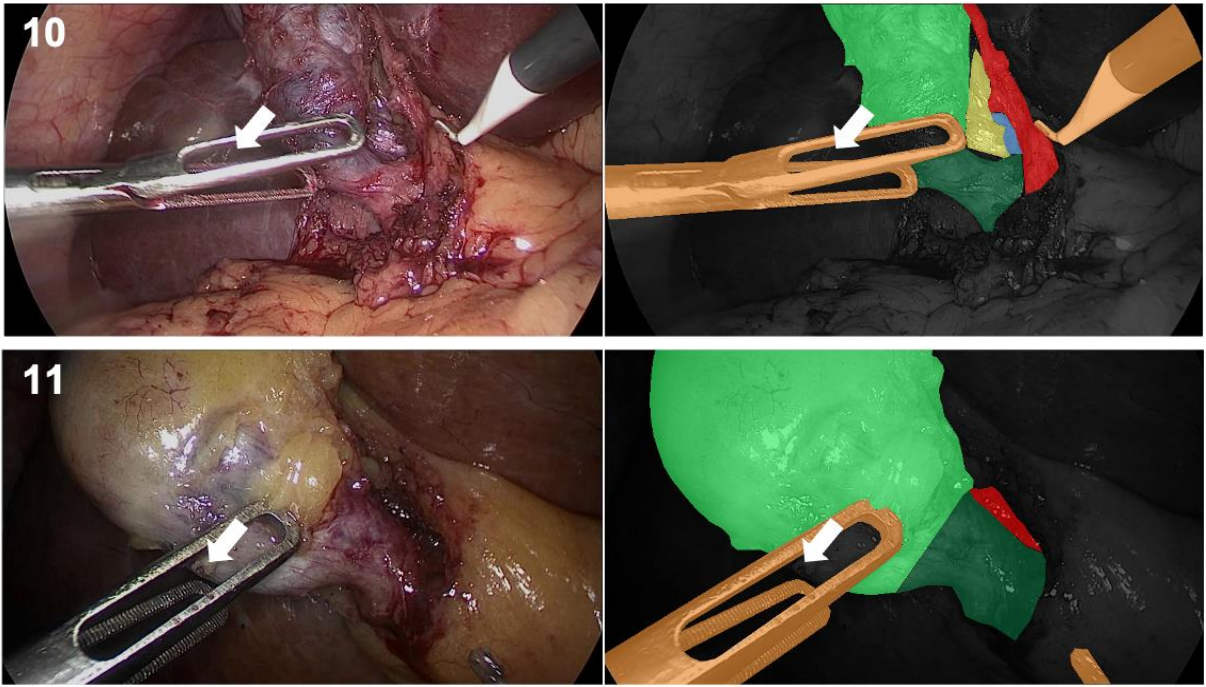

- Perform a sanity check by reviewing consistency between consecutive frames of video

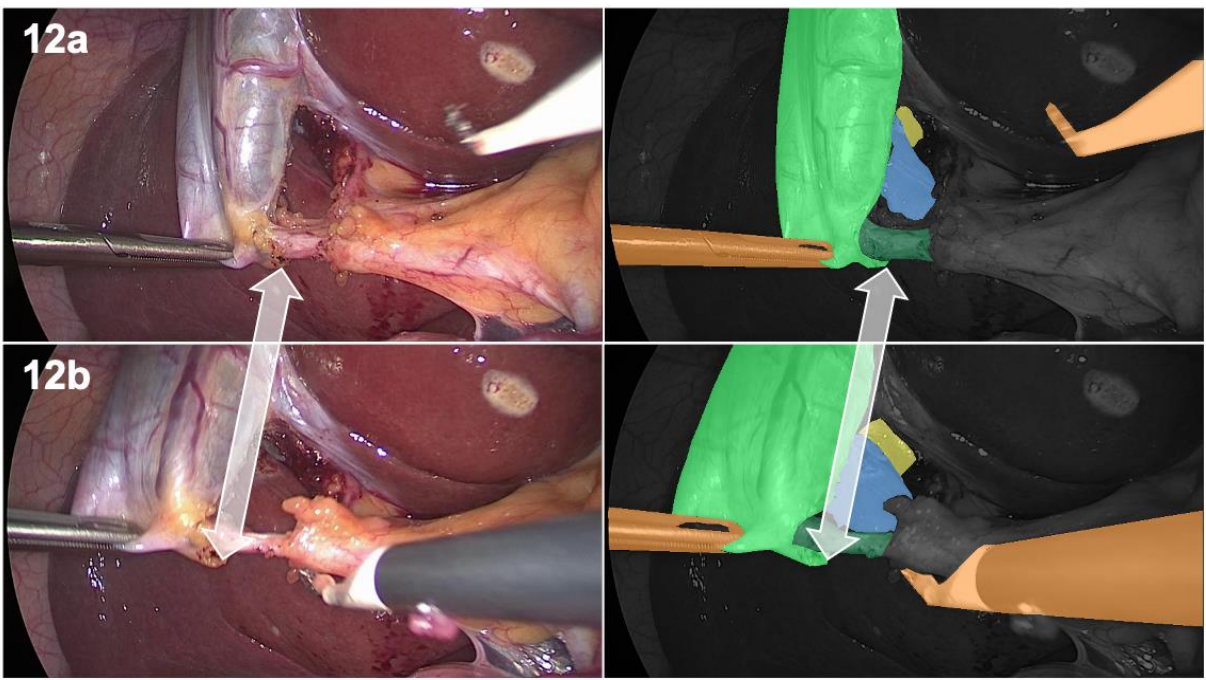

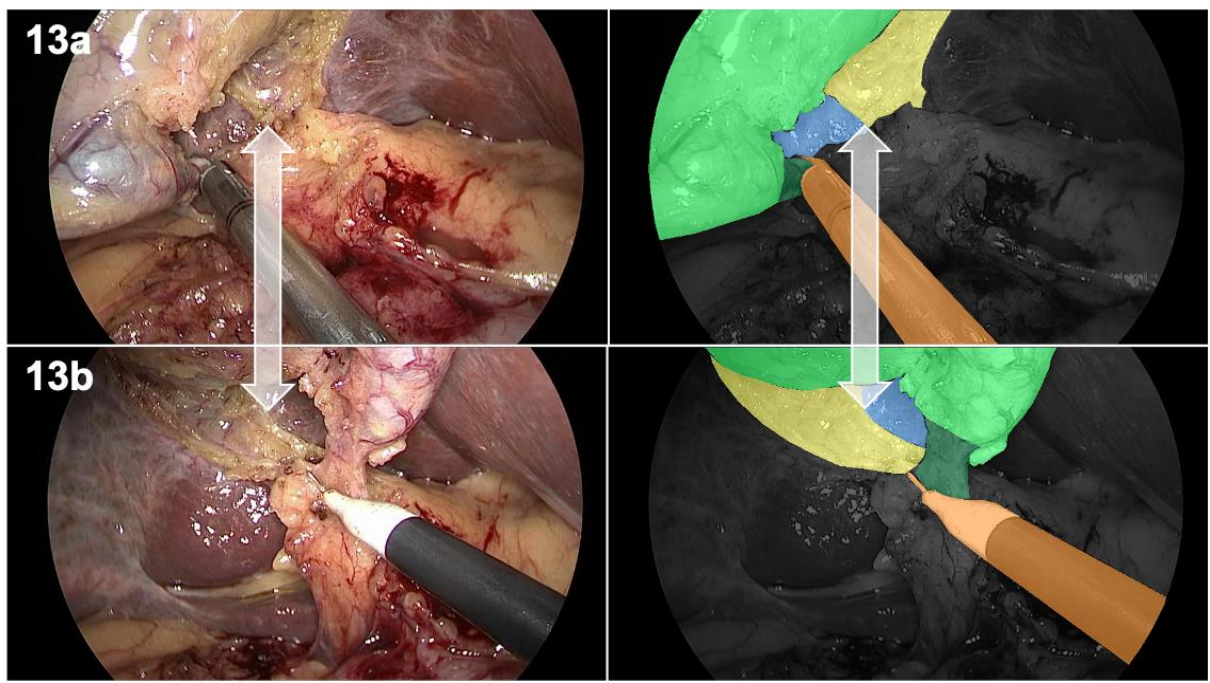

- Use the "Ignore" label (white) to mark out anatomical variants

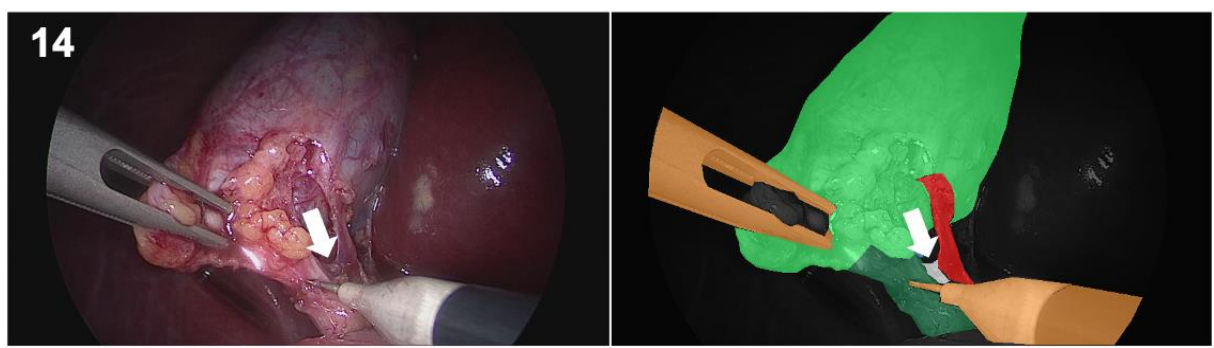